\documentclass{aip-cp}

\usepackage[numbers]{natbib}
\usepackage{rotating}
\usepackage{graphicx}

\usepackage{refstyle}

\usepackage{hyperref}

\def\ba{\begin{equation}\begin{array}{c}}
\def\ea{\end{array}\end{equation}}
\def\be{\ba\displaystyle}
\def\ee{\ea}

\newcommand{\ra}{\rangle}
\newcommand{\la}{\langle}

\renewcommand{\H}{\hat H}
\newcommand{\V}{\hat V}
\newcommand{\kF}{k_{\rm F}}

\newcommand{\mfp}{\lambda_*}

\DeclareMathAlphabet{\mathcalligra}{T1}{calligra}{m}{n}
\begin{document}

% Title portion
\title{Quantum Many-Body Adiabaticity,  Topological Thouless Pump and  Driven Impurity in a One-Dimensional Quantum Fluid}

\author[aff1,aff2,aff3]{
Oleg Lychkovskiy
\corref{cor1}
%\noteref{note1,note2}
}
%\eaddress[url]{http://www.aip.org}
\author[aff4,aff5,aff6]{Oleksandr Gamayun
%\noteref{note2}
}
\author[aff5]{Vadim Cheianov
%\noteref{note2}
}
%\eaddress{anotherauthor@thisaddress.yyy}

\affil[aff1]
{
Skolkovo Institute of Science and Technology,
Skolkovo Innovation Center 3, Moscow  143026, Russia.
%Replace this text with an author's affiliation (use complete addresses). Note the use of superscript ``a)'' to indicate the author's e-mail address below. Use b), c), etc. to indicate e-mail addresses for more than 1 author.
}

\affil[aff2]
{
Steklov Mathematical Institute of Russian Academy of Sciences,
Gubkina str. 8, Moscow 119991, Russia.
}

\affil[aff3]
{
Russian Quantum Center, Novaya St. 100A, Skolkovo, Moscow
Region, 143025, Russia.
}

\affil[aff4]{
Institute for Theoretical Physics, University of Amsterdam,
Science Park 904, 1098 XH Amsterdam, The Netherlands.
}

\affil[aff5]{
Instituut-Lorentz, Universiteit Leiden,
P.O. Box 9506, 2300 RA Leiden, The Netherlands.
}

\affil[aff6]{
Bogolyubov Institute for Theoretical Physics, 14-b
Metrolohichna str., Kyiv 03680, Ukraine.
}

\corresp[cor1]{Corresponding author: O.Lychkovskiy@skoltech.ru}
%\authornote[note1]{This is an example of first authornote.}
%\authornote[note2]{This is an example of second authornote.}

\maketitle

\begin{abstract}
The quantum adiabatic theorem states that a driven system can be kept
arbitrarily close to the instantaneous eigenstate of its Hamiltonian if
the latter varies in time slowly enough. When it comes to applying the adiabatic theorem in practice, the key question to be answered is how
slow {\it slowly enough} is. This question can be an intricate one, especially
for many-body systems, where the limits of slow driving and large system size may not commute. Recently we have shown how the quantum adiabaticity in many-body systems is related to the generalized orthogonality catastrophe \href{https://doi.org/10.1103/PhysRevLett.119.200401}{[Phys. Rev. Lett. 119, 200401 (2017)]}. We have proven a rigorous inequality relating these two phenomena and applied it to establish conditions for the quantized transport in the topological Thouless pump. In the present contribution we (i) review these developments and (ii) apply the inequality to establish the conditions for adiabaticity in a one-dimensional system consisting of a quantum fluid and an impurity particle pulled through the fluid by an external force. The latter analysis is vital for the correct quantitative description of the phenomenon of quasi-Bloch oscillations in a one-dimensional translation invariant impurity-fluid system.
\end{abstract}

% Head 1
\section{INTRODUCTION}

Consider a quantum system with a  Hamitonian $\H_\lambda$, where $\lambda$ is a time-dependent parameter.  For simplicity, we assume a linear dependence of $\lambda$ on time, $\lambda = \Gamma t,$ where $t$ is time and $\Gamma$ is called the driving rate. For each $\lambda$ one defines an instantaneous ground state, $\Phi_\lambda$,
which is the lowest eigenvalue solution to the Schr\"odinger's stationary equation,
\begin{equation}\label{eigenproblem}
\H_\lambda\, \Phi_\lambda = E_\lambda \, \Phi_\lambda.
\end{equation}
Here $E_\lambda$ is the instantaneous ground state energy. We assume that the ground state is non-degenerate for any $\lambda$.\footnote{Our consideration and results are equally applicable to other nondegenerate eigenstates. We restrict the presentation to the most important case of the ground state to simplify notations. }
The dynamics of the system is governed by the Schr\"odinger  equation, which can be written in a convenient rescaled form,
\begin{equation}\label{Schrodinger equation}
i\,\Gamma\, \frac{\partial}{\partial \lambda} \Psi_\lambda = \H_{\lambda}\,\Psi_\lambda.
\end{equation}
Here $\Psi_\lambda$ is the state vector of the system, which depends on time through the time-dependent parameter $\lambda$.  Initially the system is prepared in the instantaneous ground state:
\begin{equation}\label{initial condition}
 \Psi_0= \Phi_0.
\end{equation}

The evolution is called adiabatic as long as the state of the system, $\Psi_\lambda$, stays close to the instantaneous ground state,  $\Phi_\lambda$. The celebrated Quantum Adiabatic Theorem (QAT) \cite{Born1926,born1928beweis} states that for  however small $\epsilon>0$ and arbitrary point in the parameter space, $\lambda$,
%in the range of the function $\lambda(t)$
there exists $\Gamma$ small enough that
\be\label{adiabatic allowance}
1-\mathcal F(\lambda)<\epsilon,
\ee
where the adiabatic fidelity,
\begin{equation}\label{fidelity}
 \mathcal F(\lambda) \equiv \left | \langle \Phi_\lambda
 \vert \Psi_\lambda \rangle\right|^2,
\end{equation}
quantifies how close $\Psi_\lambda$ and $\Phi_\lambda$ are.

The QAT as presented above is a typical existence theorem. To make this theorem practical  one  usually needs to  understand how slow {\it slowly enough} is. In other words, one would like to estimate the maximum allowed $\Gamma$ for given $\lambda$ and $\epsilon$. Upper (lower) bounds on such a threshold value of $\Gamma$ are known as  necessary (sufficient) adiabatic conditions. A variety of sufficient adiabatic conditions can be constructed as byproducts of the proof of the QAT (see, e.g. \cite{jansen2006bounds}). Unfortunately, these conditions typically contain operator norms of the time derivatives of the Hamiltonian. This fact limits their applicability for continuous systems (where these operator norms are often infinite) and for many-body systems (where these operator norms, even when finite, grow very rapidly with the system size).

Alternatively, one can be interested in another meaningful question:  For a given driving rate $\Gamma$, how far in the parameter space the system can evolve whilst maintaining adiabaticity with a given allowance\footnote{in the sense of Eq. \ref{adiabatic allowance}} $\epsilon$? The answer to this question, for a given $\epsilon$, can be encoded in the {\it adiabatic mean free path}, $\mfp$. To simplify the notations we will discuss the adiabatic mean free path with respect to the fixed $\epsilon=1-1/e$, in which case it is given, by definition, by the solution of the equation\footnote{To be more exact, the mean free path is given by the smallest positive solution of Eq. (\ref{mfp}).}
\be\label{mfp}
\mathcal F(\lambda_*)=1/e.
\ee
From general considerations one expects that for gapless many-body systems $\lambda_*$ vanishes in the thermodynamic limit (TL) $N \rightarrow \infty,$ $L\rightarrow\infty$ with $n \equiv N/L^d={\rm const}$, where $N$ is the number of particles, $L$ is the linear size of the system and $d=1,2,3$ is the dimensionality of the system (see e.g.  \cite{balian2007microphysics}).\footnote{In what follows we slightly abuse the notations and denote the thermodynamic limit by $N \rightarrow \infty,$ without mentioning explicitly that  $L\rightarrow\infty$ and $n \equiv N/L^d={\rm const}$.} The  latter statement, however, has been explicitly verified only for a limited number of many-body systems \cite{polkovnikov2008breakdown,altland2008many,altland2009nonadiabaticity}. Moreover, its validity for a driven one-dimensional impurity-fluid system has been recently questioned \cite{schecter2014comment}.
%We note that ... The fidelity, (\ref{Fdef}, is a monotonically decreasing function of time, therefore one can define the adiabatic mean free path $\mfp$ as the solution to $\mathcal F(\lambda_*)=1/e$

In the recent paper \cite{lychkovskiy2017time} we have shown how the quantum adiabaticity in many-body systems can be quantitatively related to a phenomenon of a genuinely many-body origin -- orthogonality catastrophe. This relation have been used to establish a necessary condition for quantum adiabaticity and express the mean free path through the orthogonality catastrophe exponent. These general results have been applied to establish conditions for quantum adiabaticity in the topological Thouless pump and to clarify the effect of the adiabaticity breakdown on the quantization of the charge transport. In the present contribution we (i) review these developments (next two sections) and (ii) apply the developed general theory to the system  consisting of a quantum fluid and an impurity particle pulled through the fluid by a constant external force (the forth section). The latter analysis is of crucial importance for establishing conditions for quasi-Blosh oscillations in one dimension -- an intriguing phenomenon by which  an impurity particle pulled through a spatially homogeneous one-dimensional quantum fluid experiences oscillations {\it in the absence of any external periodic potential}. In particular, we resolve a dispute on whether quantum adiabaticity can be maintained in the impurity-fluid system in the thermodynamic limit \cite{schecter2014comment,Gamayun2014reply}.
%We prove that as long as the driving force does not scale with the system size, quantum adiabaticity inevitably breaks down for large systems.

\section{ADIABATICITY AND ORTHOGONALITY CATASTROPHE}

Orthogonality catastrophe is a genuine multiparticle phenomenon by which the ground states of Hamiltonians $\H_0$ and $\H_\lambda$ can be nearly orthogonal for large system sizes despite $\lambda$ is small \cite{anderson1967infrared}. To quantify this phenomenon we introduce the orthogonality overlap
\be
\mathcal C(\lambda) \equiv
\vert \langle  \Phi_{\lambda}
 \vert \Phi_{0}\rangle \vert^2 = e^{-C_N \lambda^2+r(N, \lambda)},
\ee
and say that the orthogonality catastrophe takes place whenever $C_N\to \infty$ in the thermodynamic limit.
Here $(-C_N \lambda^2)$ is the leading term of $\log \mathcal C(\lambda)$ in the thermodynamic limit in the sense that the remainder $r(N, \lambda)$ satisfies $r(N, C_N^{-1/2}) = o(1), \quad N\to \infty.$

 Note that the scaling of $C_N$ can be very different depending on the nature of driving and on whether the system is gapless or gapped in the TL.

The central result of Ref. \cite{lychkovskiy2017time} is an inequality which binds the adiabatic fidelity,  $\mathcal F(\lambda)$, and the orthogonality overlap, $\mathcal C(\lambda)$. Here we quote this inequality for an important special form of the Hamiltonian,
\be
\H_\lambda=\H_0+\lambda \V.
\ee
In this case the inequality reads \cite{lychkovskiy2017time}
\begin{equation}
\label{inequality}
|{\cal F}(\lambda)-{\cal C}(\lambda)|  \leq
\frac{\lambda^2}{2\Gamma}\,\delta V_N,
\end{equation}
where $\delta V_N$ is the uncertainty of the driving term $\V$,
\be
\delta V_N \equiv \sqrt{\langle \hat V^2 \rangle_0
- \langle   \hat V \rangle_0^2}.
\ee

The subscript "$N$" in $\delta V_N$ emphasizes that this quantity can (and in certain cases does) diverge in the thermodynamic limit. However, for a broad class of systems the orthogonality exponent, $C_N$, diverges faster:
 \begin{equation}
  \frac{\delta V_N}{C_N} \to 0, \qquad N\to \infty.
  \label{cond2}
\end{equation}
%We assume this in what follows.
 This fact has profound implications. Indeed, assume first that the driving rate, $\Gamma$, does not scale with the system size. Then
there is a timescale on which the right hand side of the inequality (\ref{inequality}) is still small and, at the same time, the orthogonality overlap has already vanished. As a consequence, the adiabaticity, which is tied to the orthogonality catastrophe by the inequality (\ref{inequality}), inevitably breaks down at this timescale.
This allows one to find the adiabatic mean free path which reads \cite{lychkovskiy2017time}
\begin{equation}
 \mfp = C_N^{-1/2}.
 \label{mainresult}
\end{equation}

The physical scenario behind this adiabaticity breakdown has been qualitatively described in Ref.~\cite{allez2012eigenvector}:
In a many-body system $\Psi_\lambda$ departs from $\Psi_0=\Phi_0$ much slower than $\Phi_\lambda$ from $\Phi_0$,  as a result $\mathcal F(\lambda)\simeq \left | \langle \Phi_\lambda
 \vert \Psi_0 \rangle\right|^2=\mathcal C(\lambda)\rightarrow 0$ with $N\rightarrow0$. Ref. \cite{lychkovskiy2017time} establishes quantitative conditions when this indeed happens.

The inequality (\ref{inequality}) implies that whenever the scaling law (\ref{cond2}) applies the only way to avoid adiabaticity breakdown with increasing system size is to scale down the driving rate with the system size. Assume that $\Gamma=\Gamma_N \rightarrow 0$ with $N\rightarrow\infty$. In order for the adiabaticity to be maintained at $\lambda=\lambda_*$ with the allowance $\epsilon$ one should require that the r.h.s. of the inequality (\ref{inequality}) is greater than $1-e^{-1}-\epsilon$. This leads to the {\it necessary adiabatic condition}  \cite{lychkovskiy2017time}
\be\label{necessary condition}
\Gamma_N <  \frac{\delta V_N}{2 C_N}\frac{1}{1-e^{-1}-\epsilon}.
\ee

\section{TOPOLOGICAL THOULESS PUMP}

The topological Thouless pump \cite{thouless1983quantization} is a quantum device which transfers  charge in a quantized manner by performing a cycle in the parameter space of its Hamiltonian. In the original paper by Thouless, Ref.  \cite{thouless1983quantization}, the quantization was proved under the assumption of adiabaticity, however, the adiabatic conditions were not discussed. In Ref. \cite{lychkovskiy2017time} we have filled this gap by applying our general theory to the simplest theoretical realization of the Thouless pump -- the Rice-Mele model \cite{rice1982elementary}. Two key quantities, $C_N$ and $V_N$, where $N$ is the number of particles in the body of the pump, have been calculated. We have found that $C_N \sim N$ and $\delta V_N\sim\sqrt N$, thus validating Eq. (\ref{cond2}). We have concluded that for any given driving rate (or, equivalently,  cycle duration), the adiabaticity breaks down for a sufficiently large pump. Alternatively, in order to have a chance to maintain adiabaticity for larger and larger systems, one needs to decrease the driving rate at least as fast as $1/\sqrt N$.

Quite remarkably, the considered model of the pump has an energy gap between the many-body ground state and the first excited state which does not vanish in the thermodynamic limit in any point of the cycle. This fact illustrates a failure of the conventional wisdom which asserts that the maximal driving rate appropriate for maintaining adiabaticity scales with the system size in the same way as the gap. This conventional wisdom is widely used but, in general, wrong. This is highlighted by the present example, where a finite gap fails to protect adiabaticity in the thermodynamic limit.

While the true many-body adiabaticity defined according to Eq. (\ref{adiabatic allowance}) is sufficient for the quantized transport \cite{thouless1983quantization}, whether it is necessary has been  remaining an open question. We have addressed this question in Ref.  \cite{lychkovskiy2017time}, with a surprising results. It appears that, in fact, two distinct modes of operation of the pump should be considered separately.

The first mode is a continuous one, when the pump performs one cycle after another, approaching a stationary state. We have verified that the true many-body adiabaticity is mandatory for quantization of the transferred charge per cycle in this mode.

The second mode can be called a transient one: One measures the  transferred charge immediately after a single cycle is completed, and then initiates the pump back in its ground state (such initialization requires some sort of external cooling). In this mode the quantization is present even when the many-body adiabaticity has gone completely. It seems plausible that  in this case a less stringent notion of a local adiabaticity \cite{bachmann2016adiabatic} can provide an adequate condition for the transferred charge quantization.

\section{DRIVEN IMPURITY IN A ONE-DIMENSIONAL FLUID}

In the present section we apply the general theory developed in \cite{lychkovskiy2017time} and reviewed above to a one-dimensional impurity-fluid system. This system can feature a phenomenon of quasi-Bloch oscillations by which an impurity particle pulled with a constant force through a one-dimensional quantum fluid experiences periodic oscillations of its velocity. Note that in contrast to the ordinary Bloch oscillations, there is no external periodic potential here and the system is translation invariant. This intriguing phenomenon has been predicted in \cite{Gangardt2009} and observed experimentally in \cite{meinert2016bloch}. While the existence of this phenomenon is beyond any doubt, the quantitative conditions for its emergence are a matter of controversy and debate \cite{schecter2014comment,Gamayun2014reply,schecter2012dynamics,schecter2012critical,Gamayun2014,Gamayun2014keldysh,Lychkovskiy2014,schecter2016quantum}.
A key issue in this debate is whether the many-body adiabaticity can be maintained  for a small but finite driving force in the thermodynamic limit \cite{schecter2014comment,Gamayun2014reply}.

Here we address this issue for a particular impurity-fluid model. This model consists of $N$ fermions and a single impurity particle with a mass $m$ equal to the mass of a fermion. The force $F$ is applied to the impurity. Fermions do not interact with each other but couple to the impurity via the repulsive contact potential. The Hamiltonian reads
\begin{equation}\label{H McGuire}
\H_Q=\frac{(- i \, \partial_X+Q)^2}{2m}-\sum_{j=1}^{N}
\frac{1}{2m}\partial_{x_j}^2+g \sum_{j=1}^{N}  \delta(x_j-X),
\end{equation}
where $Q=F t$ is the impulse of the force,
 $X$ and $x_j$  are the coordinates of the impurity and the $j$'th fermion
respectively  and $g>0$ is the impurity-fluid coupling. The role of the time-dependent parameter is played by the  dimensionless impulse $Q/\kF$, where  $\kF\equiv \pi n$ is the Fermi momentum.

For a fixed $Q$ the model (\ref{H McGuire}) is integrable as shown by McGuire \cite{mcguire1965interacting}. In fact, this model is one of the simplest models solvable via the Bethe ansatz: Its eigenfunction can be expressed through $(N+1)\times(N+1)$ Slatter-like determinants \cite{recher2012hardcore,mathy2012quantum}. For this reason it has been possible to obtain a wealth of analytical results and to gain a number of deep insights into the physics of the model \cite{recher2012hardcore,mathy2012quantum,castella1993exact,Burovski2013,gamayun2015impurity,gamayun2016time,gamayun2017quench,gamayun2018nk}. Although this model is a special case of the Yang-Gaudin model \cite{yang1967some,gaudin1967systeme}, it might deserve a separate name -- McGuire model -- due to its conceptual importance.

Thanks to the integrability, we are able to calculate $C_N$ and $\delta V_N$ for the model (\ref{H McGuire}) explicitly and to evaluate Eq.~(\ref{mainresult}) and Eq. (\ref{necessary condition}). We find that $C_N\sim \log N$ and $\delta V_N=O(1)$ in the thermodynamic limit.  The details of calculations will be presented elsewhere \cite{gamayun2018nk}, here we report the results.

The adiabatic mean free path reads
\begin{equation}
Q_*= \frac{\sqrt 2 b}{\sqrt{ \log N}}\,\kF,
\end{equation}
\be
b \equiv \left(\frac1z+z\right)\arctan z, ~~~ z\equiv\frac{2 \kF}{m g}.
\ee
One can see that in the thermodynamic limit of $N\rightarrow\infty$  the adiabatic mean free path, $Q_*$, vanishes and thus the adiabaticity breaks down for any finite value of the force.
%The dimensionless constant $b$ is a dimensionless constant dependent on $g$, $m$ and $\kF$ ($b \simeq 1$ for $mg\gtrsim\kF$, see \cite{supplementary}).
%This proves that for however small but fixed force the many-body adiabaticity breaks down with increasing system size.
%In other words,
%One can see that for a fixed force the adiabatic mean free path, $Q_*$, becomes shorter than one period of Bloch oscillations in a sufficiently large system.

If one allows the force to be dependent on the system size, $F=F_N$, one gets from Eq. (\ref{necessary condition}) the
 necessary adiabatic condition
\begin{equation}\label{necessary McGuire}
F_N\leq\frac1{\log N}\,b^2\,m^{-1}\,\kF^2\,\delta P \, \frac{1}{1-e^{-1}-\epsilon},
\end{equation}
where $\delta P$ is a quantum uncertainty of the impurity momentum,
\be\label{delta P BA}
\delta P \equiv \sqrt{ \la\hat P^2\ra_0 - \la\hat P\ra_0^2 }=\sqrt{\frac{(1-\frac2\pi \arctan z)(z-\arctan z)}{z^2\,\arctan z}}~\kF.
\ee

\section{SUMMARY}

To summarize, we have reviewed the formalism developed in Ref. \cite{lychkovskiy2017time}, which relates the adiabaticity in many-body systems to the orthogonality catastrophe, and its application to the topological quantized pumping. We have  also applied this formalism to a particular one-dimensional impurity-fluid model in which a force applied to the impurity pulls the latter through the fluid. We have found the adiabatic mean free path and established a necessary adiabatic condition for this model. As a corollary, we have proven that the adiabaticity breaks down in the thermodynamic limit  for any finite force. This way we have resolved a controversy of key importance for establishing the conditions for quasi-Bloch oscillations \cite{schecter2014comment,Gamayun2014reply}.

%We conclude by noting that an effective low-energy model can be employed to study quantum adiabaticity in a  treat a generla impurity-fluid system in a general case  the driven   general one-dimensional impurity-fluid can be treated within the formalism of Ref. \cite{lychkovskiy2017time}  an effective model

% Sections that will go in second font

% Acknowledgement
\section{ACKNOWLEDGMENTS}
The authors are grateful to P. Ostrovsky, S. Kettemann,
I. Lerner, G. Shlyapnikov, Y. Gefen, S. Nakajima, M. Schecter and M.~Troyer for fruitful discussions and
useful comments. OL acknowledges the support from the Russian Foundation for Basic Research
under Grant No. 16-32-00669. The work of OG was partially supported by Project 1/30-2015  ``Dynamics and topological structures in Bose-Einstein condensates of ultracold gases''  of the KNU  Branch Target Training at the NAS of Ukraine.

% References

\nocite{*}
\bibliographystyle{aipnum-cp}%
%\bibliography{C:/D/Work/QM/Bibs/1D,C:/D/Work/QM/Bibs/LZ_and_adiabaticity,C:/D/Work/QM/Bibs/orthogonality_catastrophe,C:/D/Work/QM/Bibs/QSL,C:/D/Work/QM/Bibs/QIP}
\bibliography{ICQT2017}

\end{document}